\documentclass[aps,twocolumn,floatfix]{revtex4-1}
\usepackage{bm}
\usepackage{dcolumn}
\usepackage{graphicx}
\usepackage{amsmath}
\usepackage{amssymb} 
\usepackage{amsfonts}
\usepackage{float}
\usepackage{xcolor}
\usepackage[capitalise]{cleveref}

\begin{document}
\title{Photoinduced dc Hall current in few-layer black phosphorus with a gate-tunable Floquet gap} 
%\author{Students}
\author{Taehun Kim}
\thanks{These authors contributed equally to this work.}
\author{Hansol Kim}
\thanks{These authors contributed equally to this work.}
\author{Dongeun Kim}
\author{Hongki Min}
\email{hmin@snu.ac.kr}
\affiliation{Department of Physics and Astronomy, Seoul National University, Seoul 08826, Republic of Korea}
\date{\today}

\begin{abstract}
We theoretically explore Floquet engineering in few-layer black phosphorus (fBP) under time-periodic driving. Motivated by the ability of circularly polarized light to induce nontrivial topological states at Dirac nodes, we investigate the emergence of a photoinduced dc Hall effect in the Dirac semimetal phase of fBP. Starting from a low-energy continuum model, we derive the effective Floquet Hamiltonian and analytically calculate the Berry curvature, demonstrating the opening of a topological gap. We also perform lattice-model calculations incorporating a self-consistent Hartree method to compute Floquet band structures and dc Hall conductivity under a perpendicular electric field. Our results reveal that the dc Hall current in fBP can be effectively tuned via a periodic driving field and electrostatic gating.

% By applying time-periodic circularly polarized light to few-layer black phosphorus (fBP) in the Dirac semimetal phase, a Floquet gap opens at the Dirac nodes, leading to a photoinduced dc Hall current that can be tuned by a perpendicular electric field.
\end{abstract}

\maketitle

\section{Introduction}

Floquet engineering has garnered significant attention for its potential to create topological states via time-periodic potentials. When the Hamiltonian is periodically driven, the system is described by Floquet states, which serve as the time-periodic analogs of Bloch states in spatially periodic systems \cite{Shirley1965, Dunlap1986, Wang2013, Mahmood2016, Sambe1973}. By coupling electronic states with photons in a time-periodic manner, quantum states in solid-state systems can be dynamically controlled, leading to the emergence of exotic quantum properties \cite{Fausti2011, Oka2019, Rudner2020, Broers2021, Liu2025}. In particular, the continuous irradiation of intense circularly polarized light can break time-reversal symmetry \cite{Claassen2017, Holder2020, Andberger2024}, facilitating the transition from a topologically trivial phase to a nontrivial one, thus generating a Floquet Chern insulator under nonequilibrium conditions \cite{Wang2018, Lindner2011, Tahir2014, Sengupta2016, Chen2021, Kong2022, Cao2024, Rechtsman2013}.

The photovoltaic Hall effect, observed in materials under an intense light field \cite{Hall1879, Haldane1988, Nagaosa2010, Liu2016, Kotegawa2023, Matsuda2020, Nguyen2021, Berdakin2021}, was first theoretically proposed for graphene \cite{Oka2009, Kitagawa2011, Perez-Piskunow2014, Zhai2014, Sato2019, FoaTorres2014, Pena2024} and later experimentally verified \cite{McIver2020}, demonstrating that a circularly polarized ac field under a dc bias can induce a photovoltaic dc Hall current. Similarly, in another two-dimensional (2D) material, black phosphorene, a topological phase transition induced by circularly polarized light has been studied theoretically \cite{Liu2018, Kang2020, Sengupta2021}. 
Experimentally, momentum-resolved band renormalization upon near-resonance pumping \cite{Zhou2023a} and below-gap pumping \cite{Zhou2023b} has been demonstrated in semiconducting bulk black phosphorus, exhibiting a strong pump polarization dependence and giving signatures of Floquet band engineering. Furthermore, the anomalous Hall, Nernst, and thermal Hall effects in semiconducting black phosphorus thin films irradiated by off-resonant circularly polarized light have been investigated using the continuum model \cite{Sengupta2021, Zhou2024}. 
%In semiconducting bulk black phosphorus, it was demonstrated that selective hybridization occurs along the armchair direction but not in the zigzag direction when a linearly polarized light is applied \cite{Zhou2023a,Zhou2023b,Zhou2024}.

\begin{figure}[htb]
\includegraphics[width=1.0\linewidth]{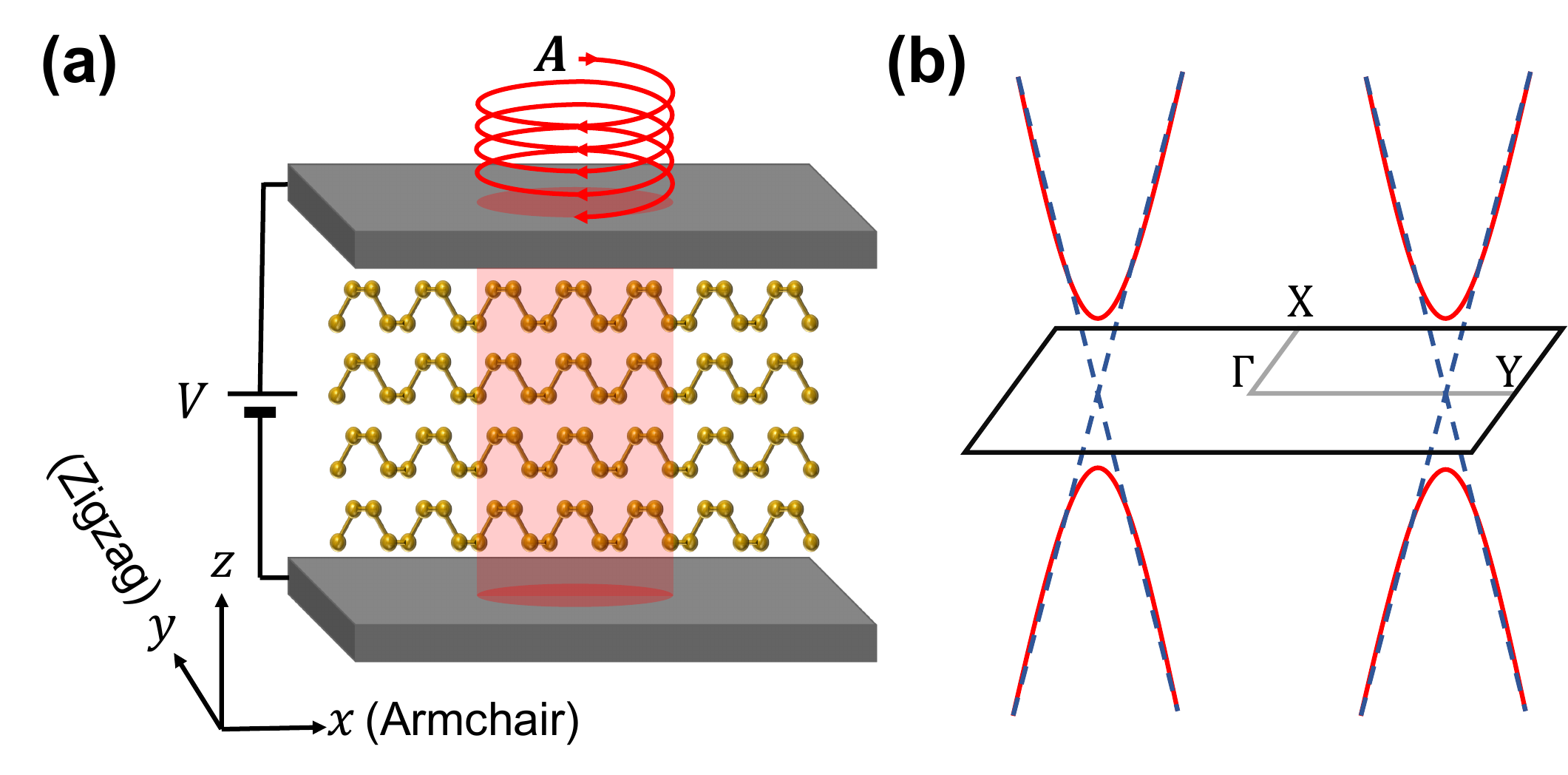}
\caption{
 Schematic illustration of (a) fBP irradiated by circularly polarized light with a voltage difference between the top and bottom layers and (b) the Floquet states in the Dirac semimetal phase of fBP in the absence (dashed lines) and presence (solid lines) of circularly polarized light. Here, $\rm{X}$ and $\rm{Y}$ represent the armchair and zigzag directions in momentum space, respectively.
} 
\label{fig:fig1}
\end{figure}

In few-layer black phosphorus (fBP), it has been demonstrated that external perturbations, such as a perpendicular electric field, can close the band gap and even induce a band inversion \cite{Liu2015, Yuan2016, Doh2017}, resulting in a transition from an insulating phase with a finite energy gap to a Dirac semimetal phase characterized by two separate Dirac nodes \cite{Low2014, Bradlyn2016, Jang2019, Saha2016, Chen2018}. The two Dirac nodes in the Dirac semimetal phase have opposite signs of Berry phase \cite{Xiao2007, Zhang2005, Delplace2013}, and in particular under optical pumping, a quantized Hall response has been demonstrated without \cite{Oka2009, Kitagawa2011}, and with spin-orbit coupling  \cite{Pena2024} in graphene. Furthermore, in  $\mathrm{Bi}_2\mathrm{Se}_3$, a 3D Dirac semimetal, optical pumping induces a gap in surface states, leading to a quantized Hall response \cite{Fang2023}. These findings suggest the possibility that a dc Hall current can be induced by circularly polarized light, and tuned by applying a perpendicular electric field.

In this paper, we theoretically investigate the effect of time-periodic potentials on fBP in the Dirac semimetal phase as shown in Fig.~\ref{fig:fig1}. First, using a continuum model \cite{Pereira2015, deSousa2017}, we analytically derive the effective Floquet Hamiltonian \cite{Rahav2003, Goldman2014} and its Berry curvature, demonstrating that a topological gap opens dynamically. Then, using a lattice model \cite{Rudenko2014, Rudenko2015, deSousa2017}, we calculate the Floquet energy bands and Hall conductivity while varying a perpendicular electric field, demonstrating the possibility of a photoinduced dc Hall current with a gate-tunable Floquet gap.

The rest of the paper is organized as follows. In Sec. II, we first describe Floquet theory, then apply it to the continuum model within the perturbation formalism and to the lattice model incorporating a self-consistent Hartree method in fBP. In Sec. III, we present the Floquet gaps and Berry curvature of charge-neutral fBP under the irradiation of circularly polarized light. Finally, in Sec. IV, we conclude with discussions on the effect of the number of layers, the feasibility of experimentally realizing the Floquet states in fBP, and the dependence of the Floquet gap on both the perpendicular electric field and the amplitude of the time-periodic potential.

\section{Model}

\subsection{Floquet theory}

% In this section, we summarize linear response theory for a time-periodically driven system by considering the Floquet quasi modes. Under the circularly polarized light, the Schr\"odinger equation is recast for the time-periodic Hamiltonian $\mathcal H(t)$ with period $T=2\pi/\Omega$ as
% \begin{equation}\label{Time-dep Schrodinger}
%     (\mathcal H(t) - i\hbar \partial_t) \left| \phi_\lambda(t)\right> = \varepsilon_\lambda \left| \phi_\lambda(t)\right>.
% \end{equation}
% Here, the Hamiltonian and the quasi mode can be written by $\mathcal{H}(t) = \sum_m e^{-im\Omega t} \tilde{\mathcal {H}}_m$ and $ |\phi_\lambda(t)\rangle = \sum_m e^{-im\Omega t} | \phi_\lambda^m \rangle $ where $\mathcal{\tilde H}_m = \frac{1}{T} \int_0^T dt e^{im\Omega t} \mathcal{H}(t)$ and $|\phi_\lambda^m \rangle = \frac{1}{T} \int_0^T dt e^{i m \Omega t} | \phi_\lambda(t)\rangle$, and each of which are the Fourier transformation and its coefficients, respectively. This system can be expressed in matrix form as Eq.~(\ref{eq:Floquet MatForm}) and $|\phi_\lambda^{\rm F}\rangle = (\cdots, |\phi_\lambda^{-1}\rangle, |\phi_\lambda^0\rangle, |\phi_\lambda^1\rangle, \cdots)^T$ for its eigenvector.

We consider a time-periodically driven Hamiltonian $\mathcal{H}(t) = \mathcal{H}(t+T)$ with $T=2\pi/\Omega$. According to Floquet theory, the time-dependent Schr\"odinger equation, $i\hbar\partial_t |\psi(t) \rangle = \mathcal{H}(t) |\psi(t)\rangle$, has quasi periodic solutions $|\psi(t)\rangle = \exp(-i\varepsilon_\lambda t/\hbar)|\phi_\lambda(t)\rangle$, where $|\phi_{\lambda}(t)\rangle = |\phi_\lambda(t + T) \rangle$ is the time-periodic Floquet mode with quasi energy $\varepsilon_\lambda$ restricted to the Floquet Brillouin zone $-\hbar\Omega/2 < \varepsilon_\lambda < \hbar\Omega/2$. By inserting the Fourier expansion into the Schr\"odinger equation, we obtain the time-independent Floquet Hamiltonian, which can be written in matrix form as \cite{Shirley1965, Sambe1973}
\begin{equation} \label{eq:Floquet MatForm}
    \mathcal{H}^{\rm{F}} = \begin{bmatrix}
        \ddots & \vdots & \vdots & \vdots & \reflectbox{$\ddots$}\\
        \cdots & \tilde{\mathcal{H}}_0 + \hbar\Omega & \tilde{\mathcal{H}}_{-1} & \tilde{\mathcal{H}}_{-2} & \cdots \\
        \cdots & \tilde{\mathcal{H}}_{1} & \tilde{\mathcal{H}}_0 & \tilde{\mathcal{H}}_{-1} & \cdots \\
        \cdots & \tilde{\mathcal{H}}_{2} & \tilde{\mathcal{H}}_{1} & \tilde{\mathcal{H}}_0 - \hbar\Omega & \cdots \\
        \reflectbox{$\ddots$} & \vdots & \vdots & \vdots & \ddots
\end{bmatrix},
\end{equation}
where $\tilde{\mathcal{H}}_n = \frac{1}{T}\int_0^T dt e^{in\Omega t} \mathcal{H}(t)$ is the Fourier coefficient of $\mathcal{H}(t)$. 

When a time-periodic vector potential $\bm{A}(t)$ with period $T = 2\pi/\Omega$ is applied perpendicular to the plane, the continuum Hamiltonian $H(\bm{k})$ is modified through the Peierls substitution, $H(\bm{k})\rightarrow H(\bm{k}, t) = H(\bm{k} + \frac{e}{\hbar c}\bm{A}(t))$, and thus also becomes time periodic with period $T$. The effective static Floquet Hamiltonian in the high-frequency limit can be obtained via Floquet-Magnus expansion \cite{Goldman2014}:
\begin{equation}\label{eq: Floquet_Magnus}
    H^{\rm{F}}(\bm{k}) = \tilde{H}_0(\bm{k}) + \sum_{n=1}^{\infty}\frac{[\tilde{H}_{-n}(\bm{k}),\tilde{H}_n(\bm{k})]}{n\hbar\Omega} + \mathcal{O}\left(\frac{1}{\Omega^2}\right).
\end{equation}
Throughout this paper, we set $\hbar\Omega = 16$ eV to effectively isolate the bands, enabling the use of the Floquet-Magnus approximation. In the following, Secs.~\ref{Continuum_sec} and ~\ref{lattice_sec}, we illustrate a continuum model and a lattice model of fBP under circularly polarized light.

\subsection{Continuum model}\label{Continuum_sec}
% {\bf [Introduce the effective Hamiltonian of fBP, mentioning a phase transition by external perturbations.]}
The low-energy effective Hamiltonian for fBP around the $\Gamma$ point is given by \cite{Baik2015, Jang2019, Kim2017, Rodin2014, Li2014, Rudenko2015}
\begin{equation}
\label{eq:continuum_model}
H(\bm{k})=\hbar v k_x \sigma_y+\left({1\over 2} \varepsilon_{\rm \Gamma}+\gamma {\hbar^2 k_x^2\over 2m}+{\hbar^2 k_y^2\over 2m}\right) \sigma_z,
\end{equation}
where $v$ is the effective velocity along the armchair direction ($k_x$), $m$ is the effective mass along the zigzag direction ($k_y$), $\gamma$ is a dimensionless parameter characterizing the effective mass along the armchair direction, and $\varepsilon_{\text{gap}}$ is the energy gap parameter. Note that linear terms in $k_y$ are forbidden in the off-diagonal elements of the Hamiltonian due to mirror symmetry $\mathcal{M}_y$ with respect to the $y=0$ plane \cite{Kim2017}. Interestingly, by tuning $\varepsilon_{\rm \Gamma}$ from positive to negative through external perturbations such as a perpendicular electric field, we observe a transition from an insulator ($\varepsilon_{\rm \Gamma}>0$) to a 2D Dirac semimetal ($\varepsilon_{\rm \Gamma}<0$) with two nodal points located at $\bm{k}=(0,\pm k_{\rm D})$, where $k_{\rm D}={\sqrt{m|\varepsilon_{\rm \Gamma}|}\over \hbar}$. At $\varepsilon_{\rm \Gamma}=0$, the Hamiltonian describes a semi-Dirac point with linear (quadratic) dispersion along the $k_x$ ($k_y$) direction. In the Dirac semimetal phase, the effective Hamiltonian near the two nodes is given by
\begin{equation}\label{eq: anisotropic Dirac}
    H(\bm{q}) \approx \hbar v q_x \sigma_y + \hbar v_{\rm{D}} q_y\tau_z\sigma_z,
\end{equation}
where $v_{\text{D}} = \hbar k_{\text{D}} / m$ and $\bm{q} = \bm{k} - (0, \tau_z k_{\rm{D}})^T$ with $\tau_z = \pm$ representing the two Dirac nodes. Unlike an isotropic Dirac Hamiltonian, Eq.~(\ref{eq: anisotropic Dirac}) has anisotropic velocities in the armchair and zigzag directions.
% {\bf [Derive the effective Floquet Hamiltonian near the zero energy in the high frequency limit and analyze the corresponding Berry curvature analytically. Demonstrate that a tunable photoinduced dc current can be induced in the Dirac semimetal phase. For the detailed derivation, present them in the appendix.]} 

Applying the Peierls substitution $H(\bm{k}) \rightarrow H(\bm{k}+\frac{e}{\hbar c}\bm{A}(t))$ with circularly polarized light $\bm{A}(t) = A_0(\cos\Omega t, \sin \Omega t)^T$, we obtain $\tilde{H}_n$ as
\begin{eqnarray}\label{eq: Hn's}
    \tilde{H}_n(\bm{k}) = \begin{cases}
        H(\bm{k}), \quad &n=0, \\
        \frac{\hbar v\Lambda}{2} \sigma_y + \frac{\hbar^2 \Lambda}{2m}(\gamma k_x \pm ik_y) \sigma_z, \quad & n=\pm1, \\
        (\gamma-1) \frac{\hbar^2\Lambda^2}{8m} \sigma_z, \quad &n=\pm 2, \\
        0, \quad & \text{otherwise,}
    \end{cases}
\end{eqnarray}
where the parameters in $\tilde{H}_n$ are taken to be the same as those in $\tilde{H}_0$. In order to derive an effective Floquet Hamiltonian $H^{\rm{F}}_{\rm{eff}}$, we apply the Floquet-Magnus expansion \cite{Goldman2014}. After substituting Eq.~(\ref{eq: Hn's}) into Eq.~(\ref{eq: Floquet_Magnus}), we obtain $H^{\rm{F}}_{\rm{eff}}(\bm{k}) = \bm{a}(\bm{k}) \cdot \bm{\sigma}$ where $\bm{a}(\bm{k})=(a_x(\bm{k}), a_y(\bm{k}), a_z(\bm{k}))$, and
\begin{subequations}\label{eq: hvec components}
\begin{eqnarray}
    a_x(\bm{k}) &=& -\left(\frac{ \frac{\hbar^2 \Lambda^2}{m} }{\hbar\Omega}\right)\hbar v k_y , \\
    a_y(\bm{k}) &=& \hbar vk_x, \\
    a_z(\bm{k}) &=& \frac{\varepsilon_{\rm \Gamma}}{2} + \gamma \frac{\hbar^2k_x^2}{2m} + \frac{\hbar^2k_y^2}{2m}.
\end{eqnarray}
\end{subequations}
In the Dirac phase ($\varepsilon_{\rm \Gamma}<0$), two nodal points appear at $\bm{k}=(0,\tau_z k_{\rm D})$ for $\tau_z=\pm$. Note that, except for the $n=1$ term, all commutators in Eq. (\ref{eq: Floquet_Magnus}) vanish under circularly polarized light. Under this illumination, when the Dirac nodes are present, by substituting $\bm{q} = \bm{k} - (0, \tau_z k_{\rm{D}})^{T}$, we obtain the gapped Dirac Hamiltonian given by
\begin{equation}\label{eq: HFeff near Dirac nodes}
    H_{\rm{eff}}^{\rm{F}}(\bm{q}) \approx  \hbar v q_x \sigma_y + \hbar v_{\rm{D}} q_y \tau_z\sigma_z - \frac{\Delta^2}{\hbar\Omega}\tau_z\sigma_x,
\end{equation}
where $\Delta = \hbar \sqrt{v v_{\rm{D}}} \Lambda$ and $\Lambda = \frac{e A_0}{\hbar c}$. This means that the Floquet energy gap $E_{\rm{FG}} = 2\Delta^2/\hbar\Omega$ opens at the two nodal points, and is proportional to $\sqrt{|\varepsilon_{\rm{\Gamma}}|}$, which can be tuned by applying a perpendicular electric field. 

Circularly polarized light breaks time-reversal symmetry and generates a nonvanishing Berry curvature. The Berry curvature $\Omega_{\mu\nu}^{\lambda}$ for a Hamiltonian of the form $\bm{a}\cdot\bm{\sigma}$ can be expressed as \cite{Graf2021}
\begin{eqnarray}\label{eq: Berry curvature for H_eff}
   \Omega_{\mu\nu}^\lambda = -\frac{\lambda}{2|\bm{a}|^3}\bm{a}\cdot(\partial_{\mu}\bm{a}\times \partial_{\nu}\bm{a}),
\end{eqnarray} 
where $\lambda =\pm$ is the band index. In the insulating phase ($\varepsilon_{\rm \Gamma}>0$), the low-energy states appear at the $\Gamma$ point. By substituting Eq.~(\ref{eq: hvec components}) into Eq.~(\ref{eq: Berry curvature for H_eff}) near the $\Gamma$ point, the Berry curvature is obtained as \cite{Zhou2024}
\begin{eqnarray}
    \Omega_{xy}^{\lambda} = -\Omega_{yx}^{\lambda} &=& \lambda\frac{\hbar^2 v^2}{2E_{\bm{k}}^3}  \left(\frac{\frac{\hbar^2\Lambda^2}{m}}{\hbar\Omega}\right) \nonumber \\
    &\times&
    \left[-\frac{\varepsilon_{\rm{\Gamma}}}{2}+\gamma\frac{\hbar^2 k_x^2}{2m} + \frac{\hbar^2 k_y^2}{2m}\right], 
    % \left[a_z(\bm{k}) - \varepsilon_{\rm {\Gamma}}\right],
\end{eqnarray}
where $E_{\bm{k}} = |\bm{a}(\bm{k})|$.
In the Dirac semimetal phase, near the nodal points $\bm{k} = (0, \tau_z k_{\rm{D}})$, substituting Eq.~(\ref{eq: HFeff near Dirac nodes}) into Eq.~(\ref{eq: Berry curvature for H_eff}) with $a_x(\bm{q}) = \hbar v q_x$, $a_y(\bm{q}) = \hbar v_{\rm{D}} q_y \tau_z$, and $a_z(\bm{q}) = -\frac{\Delta^2}{\hbar\Omega} \tau_z$, the Berry curvature can be evaluated as \begin{eqnarray} \label{eq: Berry curvature}\Omega_{xy}^{\lambda} = -\Omega_{yx}^{\lambda} = \lambda \frac{\hbar^2 v v_{\rm{D}}}{2 E_{\bm{q}}^3} \left( \frac{\Delta^2}{\hbar\Omega} \right), \end{eqnarray}
\noindent where $E_{\bm{q}} = |\bm{a}(\bm{q})|$. Note that the magnitude of the Berry curvature in the Dirac semimetal phase is significantly larger than in the insulating phase due to the large gap size $\varepsilon_{\rm \Gamma}$ in the latter.

The Berry curvature in Eq.~(\ref{eq: Berry curvature}) is independent of the node index $\tau_z$, while it varies with $\varepsilon_{\rm{\Gamma}}$ induced by a perpendicular electric field. This result implies that a photoinduced dc Hall current can be generated in the Dirac semimetal phase of fBP by applying time-periodic circularly polarized light.  Note that the parameters in the effective Floquet Hamiltonian vary with $\bm{A}(t)$ through coupling with photons, as will be discussed in Sec.~\ref{lattice_sec}.
For generally polarized light with different helicities and ellipticities, see the discussion in Sec. IV.

At zero temperature, the Hall conductivity for the effective Floquet Hamiltonian of charge-neutral fBP in Eq.~(\ref{eq: hvec components}) is given by the integration of the Berry curvature over momentum space as \cite{Xiao2010}
\begin{eqnarray}\label{Eq: Kubo}
    \sigma_{xy}(\omega=0) = -\frac{g_se^2}{\hbar}\int \frac{d^2k}{(2\pi)^2}\Omega_{xy}^{-}(\bm{k}).
\end{eqnarray}

\subsection{Lattice model}\label{lattice_sec}

% {\bf [Explain how the lattice model is constructed, including nearest-neighbor intralayer and interlayer hopping parameters.]}
The tight-binding lattice model for fBP can be constructed by including intralayer and interlayer hopping parameters with $3s$, $3p_x$, and $3p_z$ orbitals \cite{Rudenko2014, Rudenko2015, deSousa2017}. We consider a Hamiltonian of the following form:
\begin{equation}\label{Full lattice H}
    \mathcal{H}_{\rm{tot}}= \mathcal{H}_0 + \mathcal{H}_{\text{int}},
\end{equation}
with the noninteracting tight-binding Hamiltonian of fBP
\begin{eqnarray}\label{non-interacting lattice H}
       \mathcal{H}_0 &=& \sum_{\bm{k}, \alpha}\varepsilon_{\alpha}^{(0)}c_{\boldsymbol{k},\alpha}^{\dagger}c_{\boldsymbol{k}, \alpha} +\sum_{\boldsymbol{k}, \alpha, \beta}t_{\alpha\beta}c_{\boldsymbol{k},\alpha}^{\dagger}c_{\boldsymbol{k}, \beta}e^{i\boldsymbol{k}\cdot \boldsymbol{r}_{\alpha\beta}},
\end{eqnarray}
where $c_{\boldsymbol{k}, \alpha}^\dagger$ ($c_{\boldsymbol{k}, \alpha}$) creates (annihilates) electrons for the 2D wavevector $\boldsymbol{k}=(k_x, k_y)^T$ and state $\alpha$ (including spin, orbital, and layer degrees of freedom), $\boldsymbol{r}_{\alpha\beta} = \boldsymbol{r}_\alpha - \boldsymbol{r}_\beta$ is a 2D position vector connecting state $\alpha$ with $\beta$, and $\varepsilon_\alpha^{(0)}$ is the onsite energy. Here, $t_{\alpha\beta}$ denotes the hopping terms, including both intralayer and interlayer contributions, which depend on $\alpha$ and $\beta$. The electron-electron Coulomb interaction is given by
\begin{equation}
    \mathcal{H}_{\text{int}} = \frac{1}{2}\sum_{\boldsymbol{k, k', q}}\sum_{\alpha,\beta}V_{\alpha\beta}(q)c^{\dagger}_{\boldsymbol{k}+\boldsymbol{q}, \alpha}c^{\dagger}_{\boldsymbol{k}'-\boldsymbol{q}, \beta}c_{\boldsymbol{k'}, \beta}c_{\boldsymbol{k}, \alpha},
\end{equation}
where $V_{\alpha\beta}(q)=\frac{2\pi e^2}{\kappa q}e^{-qd_{\alpha\beta}}$ is the 2D Fourier representation of the Coulomb interaction, with the dielectric constant $\kappa$ and the perpendicular distance $d_{\alpha\beta}$ between states $\alpha$ and $\beta$. Following Ref.~\cite{Rudenko2015}, we choose ten nearest-neighbor intralayer hopping terms and four nearest-neighbor interlayer hopping terms in this work. 

% {\bf [In the presence of an interlayer potential difference, explain the layer potentials are determined within a self-consistent lattice Hartree theory.]}
When fBP is located between top and bottom metallic gates, charge carriers are redistributed, inducing an interlayer potential difference. We adopt a self-consistent lattice Hartree method \cite{Jang2019, Min2007} to determine the electrostatic potential generated by the induced charges. By applying the mean-field Hartree approximation, the interaction term can be reduced to
\begin{equation}
    \mathcal{H}_{\text{MF}} = \sum_{\boldsymbol{k}, \alpha}\varepsilon_{\alpha}^{(\text{H})}c^{\dagger}_{\boldsymbol{k}, \alpha}c_{\boldsymbol{k}, \alpha},
\end{equation}
where $\varepsilon_{\alpha}^{(\text{H})} = \sum_\beta V_{\alpha\beta}(0)n_{\beta}$ with $n_\beta=\sum_{\boldsymbol{k}}\langle c_{\boldsymbol{k}, \beta}^\dagger c_{\boldsymbol{k}, \beta}\rangle$. For simplicity, we set $\kappa = 1$, as the choice of $\kappa$ does not qualitatively affect our results. Then, the total on-site energy difference $\varepsilon^{(\text{tot})}_{\alpha\beta}$ between states $\alpha$ and $\beta$ due to the external field $E_{\rm{ext}}$ can be expressed as
\begin{equation}
    \varepsilon_{\alpha\beta}^{(\text{tot})} = \varepsilon^{(0)}_{\alpha}-\varepsilon^{(0)}_{\beta} + \varepsilon^{(\text{H})}_\alpha - \varepsilon^{(\text{H})}_{\beta} + eE_{\text{ext}}d_{\alpha\beta}.
\end{equation}

% {\bf [Explain the evolution of the parameters in the continuum model extracted from the lattice model. For the detailed results, present them in the appendix.]}
\noindent Here, given the charge densities of the top gate ($n_{\rm t}$) and bottom gate ($n_{\rm b}$), the external field is determined as $E_{\rm ext} = \frac{1}{2}(E_{\rm t} + E_{\rm b})$, where $E_{\rm t} = \frac{4\pi e}{\kappa}n_{\rm t}$ and $E_{\rm b} = - \frac{4\pi e}{\kappa} n_{\rm b}$.

\begin{figure}[htb]
\includegraphics[width=1.0\linewidth]{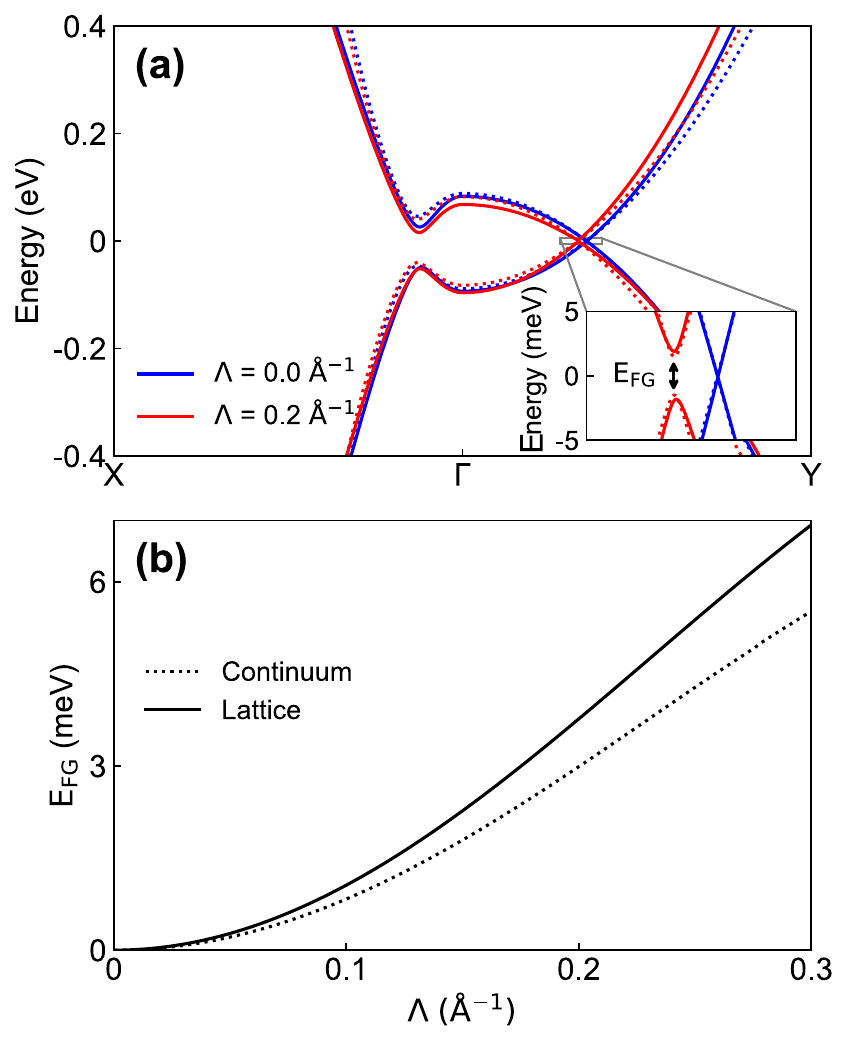} 
\caption{(a) Band structures of tetralayer BP under circularly polarized light for $\Lambda = 0.0$ \r{A}$^{-1}$ (blue) and $\Lambda = 0.2$ \r{A}$^{-1}$ (red) in the Dirac semimetal phase. (b) The Floquet gap $E_{\rm{FG}}$ in the Dirac semimetal phase as a function of $\Lambda$. In both panels, the results are presented for the lattice model (solid line) and the continuum model (dotted line) at $E_{\rm{ext}} = 0.5$ V/\r{A}. For the continuum model at $\Lambda = 0.2$ \r{A}$^{-1}$, we adopt the parameters $\varepsilon_{\rm{\Gamma}}=-0.165$ eV, $m = 1.26~m_{\rm e}$, $\gamma = 5.66$, and $\hbar v= 0.6$ eV/\r{A}, where $m_{\rm e}$ is the electron mass.
% We adopt the parameters $E_{\rm{ext}} = 0.5$ V/\r{A}, $\varepsilon_{\rm {\Gamma}}=-0.165$ eV, $m = 1.26~m_{\rm e}$, $\gamma = 5.66$, and $\hbar v= 0.6$ eV/\r{A}, where $m_{\rm e}$ is the electron mass.
} 
\label{fig:fig2}
\end{figure}

% {\bf [Construct the effective Floquet Hamiltonian in a lattice.]}
In the presence of an electromagnetic field of light generated by a vector potential $\boldsymbol{A}(t)$, the tight-binding lattice Hamiltonian is modified through the Peierls substitution, which transforms the Hamiltonian in Eq. (\ref{Full lattice H}) into
\begin{equation}\label{eq: Peierls substitution for lattice}
c^\dagger_{\alpha}c_\beta \rightarrow \exp\left[i\frac{e}{\hbar c}\boldsymbol{A}(t) \cdot\boldsymbol{r}_{\alpha\beta}\right]c^\dagger_\alpha c_\beta.
\end{equation}
Then, the Floquet Hamiltonian can be constructed as expressed in Eq.~(\ref{eq:Floquet MatForm}), and we truncate the Floquet Hamiltonian when the results are convergent. Note that the hopping terms are renormalized through coupling with photons (appearing in the form of Bessel functions in the lattice model \cite{Quelle2016}), implying that the parameters in the continuum model vary not only with the external electric field but also with circularly polarized light $\bm{A}(t) = A_0(\cos\Omega t, \sin \Omega t)^T$. Here, we assume that under a perpendicular electric field, the static Hartree potential energy remains unchanged $\varepsilon_{\alpha}^{(\rm{H})}\approx \tilde{\varepsilon}^{(\rm{H})}_{\alpha}$, where $\tilde{\varepsilon}^{(\rm{H})}_{\alpha} = \sum_{\beta} V_{\alpha\beta} \tilde{n}_{\beta}$ with $\tilde{n}_\beta = \frac{1}{T}\int_0^T dt \sum_{\bm{k}}\langle c^{\dagger}_{\bm{k}, \beta}(t)c_{\bm{k}, \beta}(t)\rangle $, since light has little effect on the overall charge distribution in the high-frequency limit \cite{Bukov2015}. The continuum model in Eq.~(\ref{eq:continuum_model}) can be obtained from the self-consistently calculated Floquet Hamiltonian by extracting the model parameters near the $\Gamma$ point, including the band renormalization due to the gate field and irradiation. The evolution of the model parameters $\varepsilon_{\rm \Gamma}$, $v$, $m$, and $\gamma$ for bilayer, trilayer, and tetralayer is presented in Appendix~\ref{sec:self_consistent_Hartree}.

\begin{figure}[htb]
\includegraphics[width=1.0\linewidth]{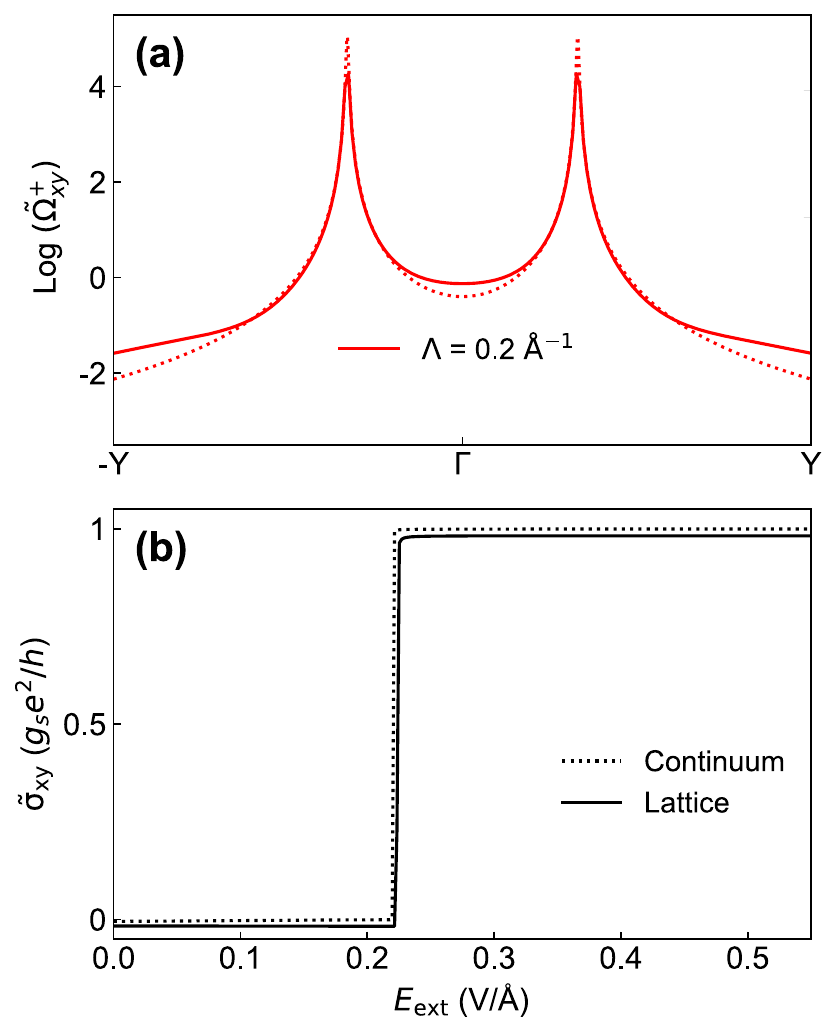}
\caption{(a) Berry curvature of tetralayer BP under circularly polarized light for $\Lambda = 0.2$ \r{A}$^{-1}$ (red) in the Dirac semimetal phase at $E_{\rm{ext}} = 0.5$ V/\r{A}. (b) The Hall conductivity $\tilde{\sigma}_{xy}$ of tetralayer BP as a function of $E_{\rm{ext}}$ at $\Lambda = 0.2$ \r{A}$^{-1}$. In both panels, the results are shown for the lattice model (solid line) and the continuum model (dotted line).
}
\label{fig:fig3}
\end{figure}

\section{Numerical results}

In this section, we present numerical results for charge-neutral tetralayer fBP under circularly polarized light. The effect of the number of layers is discussed in the discussion section. Throughout this paper, we apply a linear response theory to time-periodic quantum systems for the lattice model, as discussed in Appendix~\ref{APP_Optical_conductivity_in_Floquet}, whereas for the continuum model, we apply the conventional Kubo formula in Eq.~(\ref{Eq: Kubo}) to the effective Floquet Hamiltonian, assuming the high-frequency approximation as discussed before.

In Fig.~\ref{fig:fig2}(a), we present the band structures for the continuum and lattice models in the Dirac semimetal phase. In the continuum model, we assume that the model parameters used in $\tilde{H}_n$ are identical for all $n$, and that they are extracted from the self-consistent lattice Hartree calculation. One can see that circularly polarized light opens the Floquet band gaps at the Dirac nodes in the Dirac semimetal phase. Note that since the continuum model corresponds to the leading-order expansion of the lattice model in terms of the vector potential characterized by $\Lambda={eA_0\over\hbar c}$, the two models agree well in the limit $\Lambda \ll 1$ \r{A}$^{-1}$, while their deviation grows with increasing $\Lambda$, as shown in Fig.~\ref{fig:fig2}(b).

An anomalous velocity due to the nonvanishing Berry curvature gives rise to a Hall current. The Hall conductivity can be written as \cite{Cao2024, Kumar2020, Dabiri2025}
\begin{subequations}\label{eq: Hall conductivity}
\begin{flalign}
        \tilde{\sigma}_{xy} &= -\frac{g_se^2}{\hbar} \sum_{\lambda}\int\frac{d^2k}{(2\pi)^2}\tilde{\Omega}_{xy}^{\lambda}(\bm{k}) f_{\lambda, \bm{k}}, \\
        \tilde{\Omega}_{xy}^{\lambda}(\bm{k}) &= -2\text{Im}\left[\frac{1}{T}\int_{0}^Tdt \langle\partial_{k_x}\phi_{\lambda, \bm{k}}(t)|\partial_{k_y}\phi_{\lambda, \bm{k}}(t)\rangle \right],
\end{flalign}
\end{subequations}
\noindent where $g_s = 2$ is the spin degeneracy factor, $f_{\lambda, \bm{k}}=\Theta(-\varepsilon_{\lambda, \bm{k}})$ is the ideal occupation function of the quasi energy $\varepsilon_{\lambda, \bm{k}}$, associated with the Floquet mode $|\phi_{\lambda}(t)\rangle$ and the Floquet band index $\lambda$, and $\tilde{\Omega}_{xy}^{\lambda}(\bm{k})$ denotes the time-averaged Berry curvature in momentum space.

\begin{figure}[htb]
\includegraphics[width=1.0\linewidth]{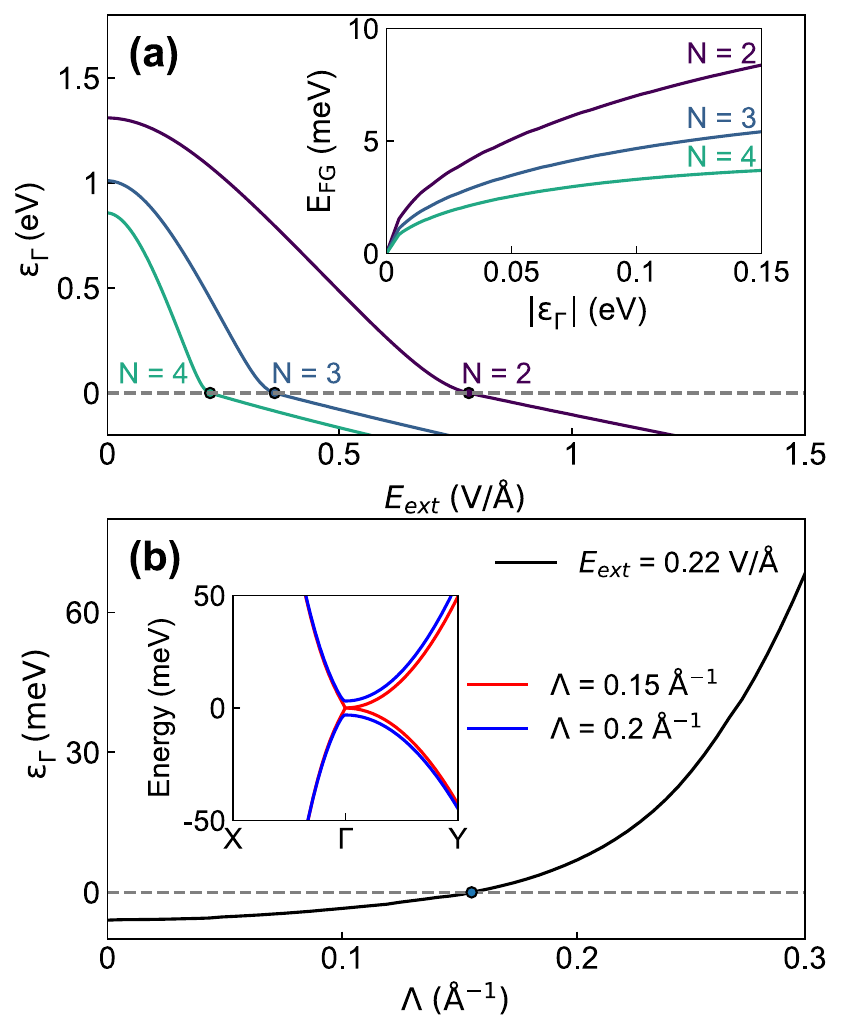}
\caption{(a) Evolution of $\varepsilon_{\rm {\Gamma}}$ as a function of $E_{\rm{ext}}$ for different layer numbers ($N = 2, 3, 4$) at $\Lambda = 0.2$ \r{A}$^{-1}$. The inset shows the corresponding Floquet gap $E_{\rm{FG}}$ as a function of $|\varepsilon_{\rm {\Gamma}}|$. (b) The evolution of $\varepsilon_{\rm {\Gamma}}$ as a function of $\Lambda$ for $N = 4$ at $E_{\rm{ext}} = 0.22$ V/\r{A}. The inset illustrates the merging of two Dirac nodes at the $\Gamma$ point when $\Lambda = 0.15$ \r{A}$^{-1}$ and the transition to an insulating phase at $\Lambda = 0.2$ \r{A}$^{-1}$. In both panels, the circles indicate the transition point between the insulating phase and the Dirac semimetal phase.}
\label{fig:fig4}
\end{figure}

Figure 3(a) describes the Berry curvature of the Floquet bands in the Dirac semimetal phase. Both the continuum and lattice models capture the Berry curvature peaks at the Dirac nodes in the Dirac semimetal phase. In the absence of irradiation, the Berry curvature vanishes due to the presence of time-reversal and inversion symmetries. When circularly polarized light is applied, however, time-reversal symmetry is broken by the driving field, and the Berry curvature becomes nonzero with the same sign at the Dirac nodes. Figure 3(b) shows the Hall conductivity as a function of $E_{\rm{ext}}$. For all values of $E_{\rm{ext}}$, when light is not applied, the Berry curvature of valence band is always zero because of time-reversal symmetry, resulting in $\tilde{\sigma}_{xy} = 0$. Under illumination, the Berry curvature becomes nonzero in both the insulating phase and the Dirac semimetal phase, although in the insulating phase, only marginal Berry curvature is obtained \cite{Zhou2024}. Note that while the Hall conductivity in the insulating phase remains negligible, it exhibits a topologically quantized value corresponding to the Chern number in the Dirac semimetal phase. The phase transition from the insulating phase to the Dirac semimetal phase occurs at $E_{\text{ext}} = 0.225$ V/\r{A} in tetralayer BP. Indeed, the Hall conductivity of the Dirac cones is quantized for charge neutral fBP and deviates from the quantized value once the conduction band becomes occupied or the valence band becomes unoccupied.

\section{Discussion}

The present analysis focuses on tetralayer BP. Varying the number of layers $N$ modifies the band structure and shifts the corresponding transition point to the Dirac semimetal phase. Figure~\ref{fig:fig4}(a) illustrates the evolution of $\varepsilon_{\rm{\Gamma}}$ as a function of $E_{\rm{ext}}$, along with the Floquet gap $E_{\rm{FG}}$ as a function of $|\varepsilon_{\rm{\Gamma}}|$ for $N = 2, 3, 4$ at $\Lambda = 0.2$ \r{A}$^{-1}$. The results demonstrate that the critical $E_{\rm{ext}}$ at which the Dirac semimetal phase emerges decreases with increasing $N$, while the value of $E_{\rm{FG}}$ for a given $|\varepsilon_{\rm{\Gamma}}|$ also decreases. This indicates that although a larger $E_{\rm{FG}}$ can be achieved at smaller $N$, it becomes more difficult to induce the Dirac semimetal phase. Nevertheless, the Hall conductivity in the Dirac semimetal phase remains quantized regardless of the number of layers, indicating that the topological nature of the system is preserved.

For typical values of $\Lambda = 0.2$ \r{A}$^{-1}$ and $E_{\rm{ext}} = 0.5$ V/\r{A} in tetralayer BP, we find that the Floquet gap reaches approximately 4 meV, placing it well within the range of current experimental detection. This energy scale is comparable to those observed in other Floquet-engineered systems \cite{Zhou2011, Karch2011}, suggesting that the predicted gap opening in fBP with a tunable perpendicular electric field could be verified using available spectroscopic techniques. Furthermore, the Floquet gap could also be increased by reducing the frequency $\Omega$ of the time-periodic potential, while additional contributions occurring at lower energy scales, such as phonons, could influence the Floquet dynamics, and the light-induced change in the charge distribution may not be neglected at low frequencies. We leave the analysis of the low-frequency behavior as an open question for future research.

We point out that the Floquet gap $E_{\rm{FG}}$ does not increase monotonically with $\Lambda$, since $\Lambda$ itself modifies the band structure. As $\Lambda$ increases, the two Dirac points move closer and eventually merge, leading to a phase transition from the Dirac semimetal phase to the insulating phase, as shown in Fig.~\ref{fig:fig4}(b).

In our calculations, we employed right-circularly polarized light. For the opposite helicity (i.e., left-circularly polarized light), the Floquet gap opens in the same manner, but the sign of the Berry curvature is reversed, resulting in a reversal of the dc Hall conductivity. In the case of elliptically polarized light, the magnitude of the Floquet gap varies quantitatively with the ellipticity. Nevertheless, the dc Hall conductivity remains quantized at charge neutrality, and its sign remains unchanged as long as the helicity is preserved. We leave the details in Appendix~\ref{sec: elliptical polarization}

In summary, we have investigated the effect of circularly polarized light on few-layer black phosphorus (fBP) in the Dirac semimetal phase, focusing on the Floquet gap at the two Dirac points and the corresponding photoinduced dc Hall current as potential experimental signatures. We constructed an effective Floquet continuum model for fBP, demonstrating that the Floquet gap scales with the gap parameter in the Dirac semimetal phase as $E_{\rm{FG}} \propto \sqrt{|\varepsilon_{\rm{\Gamma}}|}$. In addition, we developed the full lattice Floquet Hamiltonian, incorporating a self-consistent Hartree potential under irradiation, and estimated the Floquet gap by varying the external electric field and the light intensity for bilayer, trilayer, and tetralayer structures, thus establishing the feasibility of experimental realization. Furthermore, the Floquet gap is tunable by applying a perpendicular electric field through gating, highlighting the potential of fBP as a versatile platform for optoelectronic applications and paving the way toward realizing photoinduced dc Hall currents with a gate-tunable Floquet gap.

\section*{data availability}
The data that support the findings of this article are not publicly available. The data are available from the authors upon reasonable request.

\acknowledgments
This work was supported by the National Research Foundation of Korea (NRF) grants funded by the Korea government (MSIT) (Grant No. RS-2023-NR076715), the Creative-Pioneering Researchers Program through Seoul National University (SNU), and the Center for Theoretical Physics.

\appendix

\begin{figure}[htb]
\includegraphics[width=1.0\linewidth]{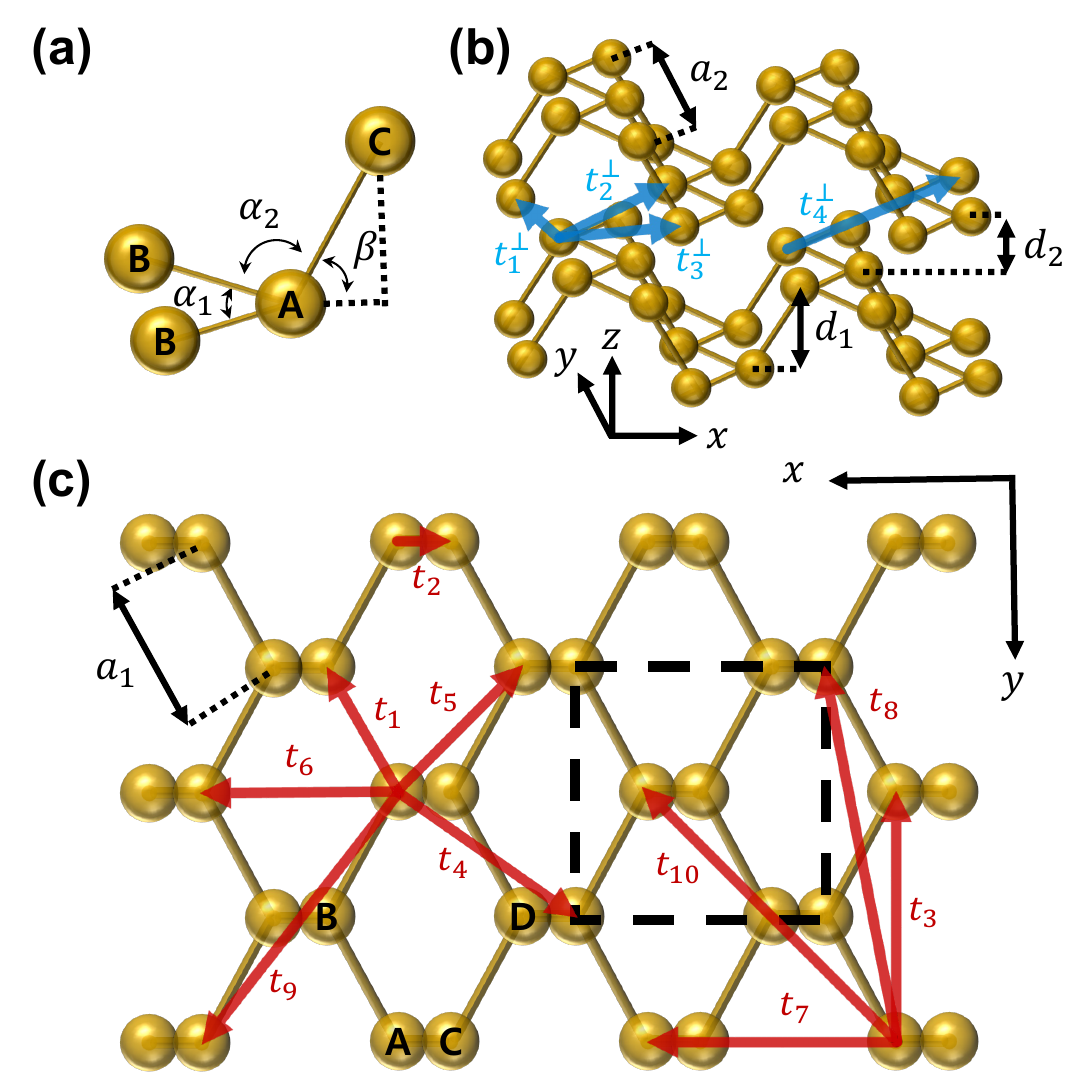}
\caption{
(a) Nearest neighbors in the phosphorene lattice, (b) atomic structure of bilayer BP, and (c) top view of monolayer BP. The out-of-plane direction is denoted by the $z$ axis, while the $x$ and $y$ axes correspond to the armchair and zigzag directions, respectively. The interatomic angles are $\alpha_1 = 96.5^{\circ}$, $\alpha_2 = 101.9^{\circ}$, and $\beta = 72^{\circ}$. The interatomic distances are $a_1 = 2.21$ \r{A} and $a_2 = 2.24$ \r{A}, and the vertical intralayer and interlayer distances are $d_1 = 2.13$ \r{A} and $d_2 = 3.17$ \r{A}, respectively. The red (blue) arrows indicate intralayer (interlayer) hopping terms $t_i$ ($t_i^\perp$).
}
\label{fig:fig5}
\end{figure}

\begin{figure*}[t]
    \centering
    \includegraphics[width=1.0\linewidth]{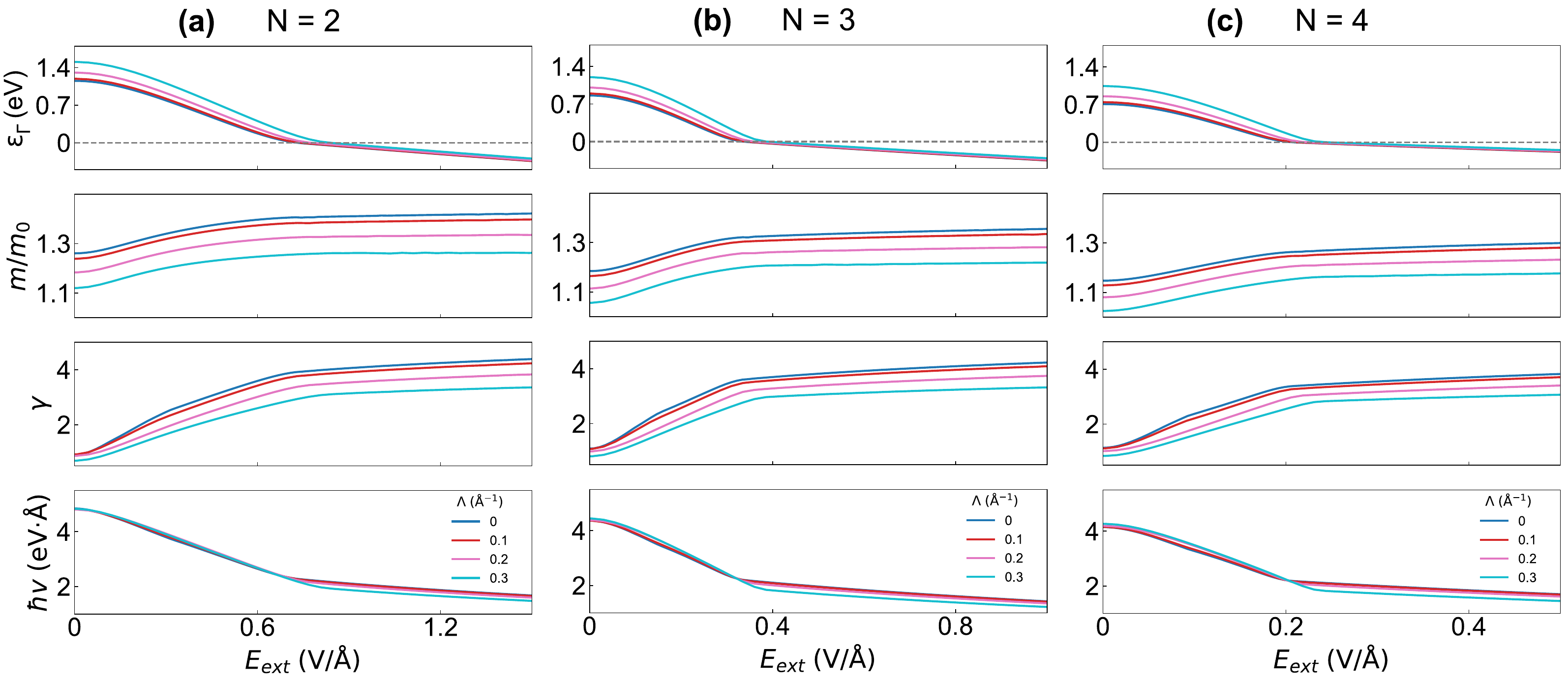}
    \caption{
    Evolution of the continuum model parameters $\varepsilon_{\rm {\Gamma}}$, $\gamma$, $m /m_{\rm e}$, and $\hbar v$ as a function of $E_{\rm{ext}}$ for (a) $N=2$, (b) $N=3$, and (c) $N=4$ using a self-consistent lattice Hartree method at $\Lambda=0, 0.1, 0.2, 0.3$ \r{A}$^{-1}$. The dashed gray line in the $\varepsilon_{\rm {\Gamma}}$ plot indicates the phase transition from the insulating phase to the Dirac semimetal phase.}
    \label{fig:fig6}
\end{figure*}

\section{Self-consistent lattice Hartree calculations}
\label{sec:self_consistent_Hartree}

%  The effects of light can be expressed in terms of the Bessel functions $J_{n}(z)$ of the first kind by using the well known Jacobi-Anger identity
% \begin{equation}
% e^{iz\sin{x}} = \sum_{n=-\infty}^{\infty}J_n(z) e^{inx}.
% \end{equation}
% By expanding these expressions, one can construct the Floquet Hamiltonian expressed in Eq.~(\ref{eq:Floquet MatForm}).

For the self-consistent lattice Hartree calculations, we consider the tight-binding lattice Hamiltonian of an $N$-layer black phosphorus system expressed in an $N \times N$ block tridiagonal matrix form as \cite{Baik2015, deSousa2017}
\begin{equation}\label{eq: tihgt-binding model for fBP}
    \begin{aligned}
        \mathcal{H}^{(N)} = \begin{bmatrix}
            \mathcal{H}_{\text{mono}} & T & & & \\
            T^\dagger & \mathcal{H}_{\text{mono}} & \ddots & & \\
             & \ddots & \ddots & T \\
             & & T^\dagger & \mathcal{H}_{\text{mono}}
        \end{bmatrix}, 
    \end{aligned}
\end{equation}
where $\mathcal{H}_{\text{mono}}$ is the Hamiltonian of monolayer black phosphorus, and $T$ is the interlayer tunneling matrix that contains the hopping terms between atomic sites in adjacent layers. These two matrices are given by
\begin{subequations}
\begin{eqnarray}
    \mathcal{H}_{\text{mono}} &=& \begin{bmatrix}
        t_{AA} & t_{AB} & t_{AD} & t_{AC} \\
        t_{AB}^* & t_{AA} & t_{AC}^* & t_{AD} \\
        t_{AD} & t_{AC} & t_{AA} & t_{AB} \\
        t_{AC}^* & t_{AD} & t_{AB}^* & t_{AA}
    \end{bmatrix}, \\
    T &=& \begin{bmatrix}
        0 & 0 & t_{AD'} & t_{AC'} \\
        0 & 0 & t_{AC'}^* & t_{AD'} \\
        0 & 0 & 0 & 0 \\
        0 & 0 & 0 & 0
    \end{bmatrix}.
\end{eqnarray}
\end{subequations}
Here, $t_{AA}, t_{AB}, t_{AC}$, and $t_{AD}$ denote intralayer hopping terms, while $t_{AC'}$ and $t_{AD'}$ represent interlayer hopping terms. Each hopping term $t_{\alpha\beta}$ is given by
\begin{subequations}\label{tight-binding lattice}
\begin{eqnarray}
    t_{AA} &=& 2t_3 \cos(2a_{1y} k_y) + 2t_7 \cos \left[ (2a_{1x} + 2a_{2x}) k_x  \right] \nonumber \\ 
    &+& 4t_{10} \cos \left[ (2a_{1x} + 2a_{2x}) k_x \right] \cos ( 2a_{1y} k_y) , \\
    t_{AB} &=& 2t_1 e^{-i a_{1x} k_x} \cos (a_{1y} k_y ) \nonumber \\
    &+& 2t_4 e^{i (a_{1x} + 2a_{2x}) k_x } \cos(a_{1y} k_y) \nonumber \\
    &+& 2t_8 e^{-i a_{1x}k_x } \cos (3a_{1y} k_y ) ,  \\
    t_{AC} &=& t_2 e^{i a_{2x} k_x } + t_6 e^{-i (2a_{1x} + a_{2x}) k_x } \nonumber \\
    &+& 2t_9 e^{-i (2a_{1x} + a_{2x})k_x} \cos(2a_{1y} k_y) , \\
    t_{AD} &=& 4t_5 \cos \left[ (a_{1x} + a_{2x}) k_x \right] \cos (a_{1y} k_y) , \\
    t_{AC'}&=& 2t_{1}^{\perp} e^{i a_{1x} k_x} \cos(a_{1y} k_y) \nonumber \\
    &+& 2t_4 ^\perp  e^{-i(2a_{1x}+a_{2x}) k_x} \cos(a_{1y} k_y) , \\
    t_{AD'}&=& 2t_2^\perp \cos \left[ (a_{1x}+a_{2x}) k_x \right] \cos (2a_{1y} k_y) \nonumber \\
    &+& 4t_3^\perp \cos \left[ (a_{1x}+a_{2x}) k_x\right] \cos(2a_{1y} k_y) ,
\end{eqnarray}
\end{subequations}
where $a_{1x} = a_1 \cos (\alpha_1/2)$, $a_{1y} = a_1 \sin (\alpha_1/2)$, and $a_{2x} = a_2 \cos\beta$. Here, $t_i$ and $t_i^\perp$ denote the ten intralayer and four interlayer hopping parameters, respectively. The definitions of the bond angles and bond lengths are shown in Fig.~\ref{fig:fig5}.

Under the periodic driving field, the Hamiltonian is modulated through the Peierls substitution as Eq.~(\ref{eq: Peierls substitution for lattice}). The effects of light can be expressed in terms of the Bessel functions $J_n(z)$ of the first kind by using the well known Jacobi-Anger identity
\begin{equation}
    e^{iz\sin{x}} = \sum_{n=-\infty}^{\infty} J_n(z) e^{inx}.
\end{equation}
Using these expressions, we construct the Floquet Hamiltonian as given in Eq.~(\ref{eq:Floquet MatForm}). We then perform a Hartree calculation, assuming a static Hartree potential based on the Floquet band structure in the high-frequency limit.
Figure \ref{fig:fig6} presents the evolution of the continuum model parameters as a function of $E_{\rm{ext}}$ under circularly polarized light. Note that the parameters depend not only on $E_{\rm{ext}}$ but also on $\Lambda$, because the hopping terms are renormalized due to coupling with photons.

\section{Optical conductivity in time-periodic quantum systems}
\label{APP_Optical_conductivity_in_Floquet}

Under the circularly polarized light, the interband optical conductivity can be given as \cite{Kumar2020, Dabiri2025}

\begin{eqnarray}
    \tilde{\sigma}_{\mu\nu}(\omega) &=& -\frac{ig_se^2}{\hbar}\sum\limits_{{\substack{\lambda, \lambda', \\\lambda \neq \lambda' } }}\sum_{n=-\infty}^{\infty}\int\frac{d^2k}{(2\pi)^2}\frac{f_{\lambda, \bm{k}}-f_{\lambda', \bm{k}}}{\varepsilon_{\lambda, \bm{k}} - \varepsilon_{\lambda', \bm{k}} + n\hbar\Omega} \nonumber \\
    &\times& \frac{M_{\mu, -n}^{\lambda\lambda'}(\bm{k})M_{\nu, n}^{\lambda'\lambda}(\bm{k})}{\hbar\omega + \varepsilon_{\lambda, \bm{k}} - \varepsilon_{\lambda', \bm{k}} + n\hbar\Omega+i\eta},
\end{eqnarray}
where $\mu, \nu = x, y$, $n$ is the photon index, $\eta$ is a positive infinitesimal number, and $M_{\mu, n}^{\lambda\lambda'}(\bm{k}) = \frac{1}{T}\int^{T}_{0} dt e^{in\Omega t}\langle \phi_{\lambda, \bm{k}}(t)|\hbar v_{\mu}|\phi_{\lambda', \bm{k}}(t)\rangle$, with the velocity operator $v_{\mu} = \frac{1}{\hbar}\frac{\partial \mathcal{H}}{\partial k_\mu}$. Note that from the Schr\"odinger equation, we have the following relation:
\begin{eqnarray}
&&\langle \phi_{\lambda, \bm{k}}(t) | \hbar v_{\mu} |\phi_{\lambda', \bm{k}}(t)\rangle 
=\partial_{k_\mu} \varepsilon_{\lambda, \bm{k}} \delta_{\lambda\lambda'} \nonumber\\
&&+ (\varepsilon_{\lambda, \bm{k}} - \varepsilon_{\lambda', \bm{k}} + i\hbar \partial_t) \langle \phi_{\lambda, \bm{k}}(t) | \partial_{k_\mu} \phi_{\lambda', \bm{k}}(t) \rangle.
\end{eqnarray}
Using the equation above, one can find that the interband optical conductivity $\tilde{\sigma}_{xy}(\omega)$ in the dc limit ($\omega \rightarrow 0$) becomes Eq.~(\ref{eq: Hall conductivity}).

\begin{figure}[htb]
\includegraphics[width=1.0\linewidth]{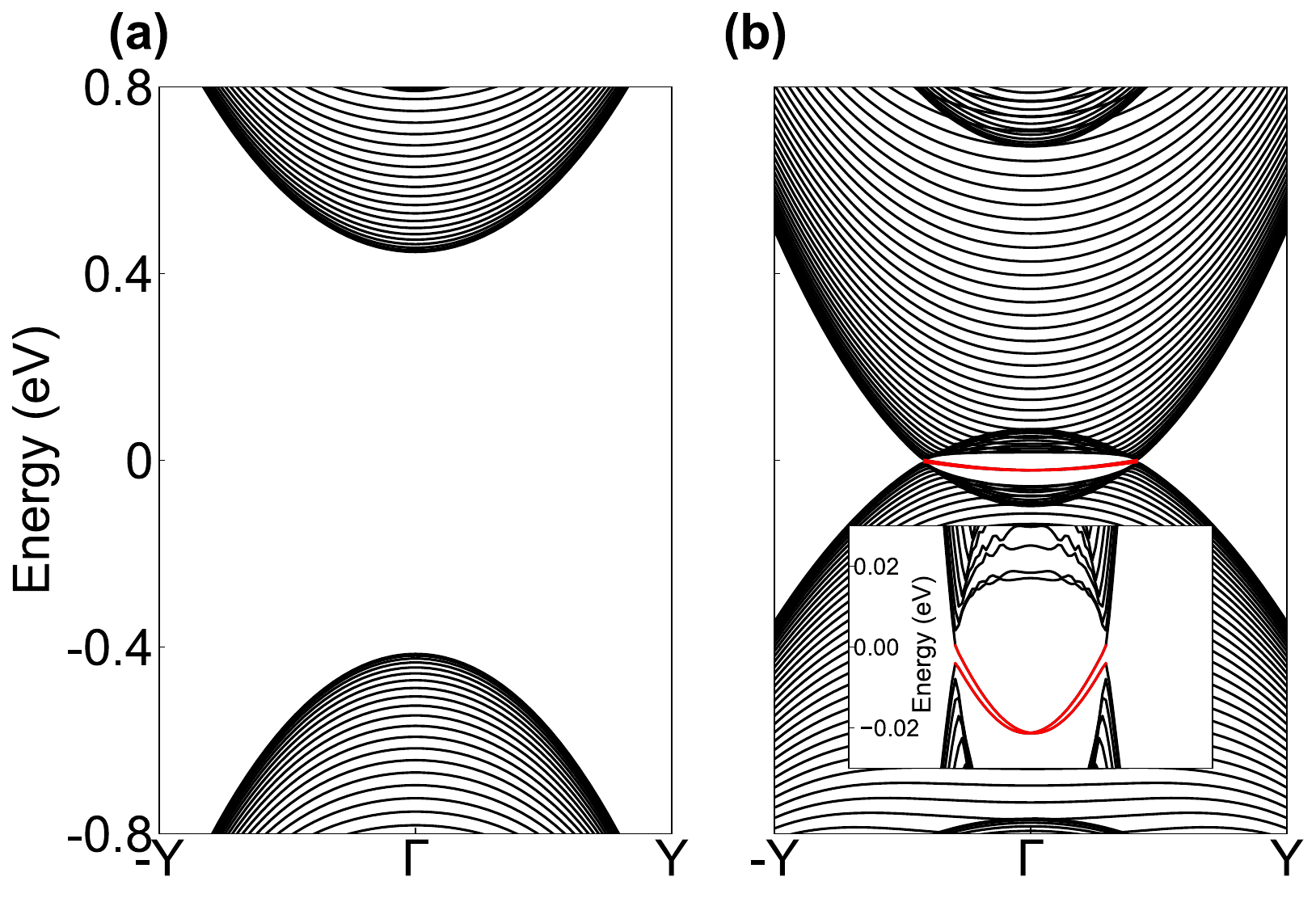}
\caption{
Band structures of bearded nanoribbons of tetralayer BP with the width $W = 100(2a_{1x} + a_{2x})$ at $\Lambda = 0.2$ \r{A}$^{-1}$ for (a) the insulating phase ($E_{\rm{ext}} = 0$ V/\r{A}) and (b) the Dirac semimetal phase ($E_{\rm{ext}}=0.5$ V/\r{A}). Edge states are indicated by red lines. The inset highlights the Floquet edge modes that connect the two Dirac nodes. 
} 
\label{fig:fig7}
\end{figure}

\section{Floquet edge states}

We use the tight-binding Hamiltonian in Eq.~(\ref{eq: tihgt-binding model for fBP}) to describe the bearded fBP nanoribbons \cite{Ezawa2014}. The nanoribbon geometry is taken to be finite along the armchair direction and infinite along the zigzag direction with a bearded edge configuration. By diagonalizing the Hamiltonian, we obtain the band structure of the bearded nanoribbons in the Dirac semimetal phase realized under irradiation. As shown in Fig.~\ref{fig:fig7}, Floquet-induced edge states, which are hallmarks of the nontrivial nature of a Chern insulator, appear between the two gapped Dirac nodes in the Dirac semimetal phase, while they are absent in the insulating phase. Note that in the absence of irradiation, where time-reversal symmetry is preserved, the edge states in the Dirac semimetal phase of bearded fBP are degenerate and connect the two gapless Dirac nodes \cite{vanMiert2017}, whereas in the presence of circularly polarized light, a Floquet gap opens at the two Dirac nodes and the edge states cross, as shown in the inset of Fig.~\ref{fig:fig7}(b).

\begin{figure}[htb]
\includegraphics[width=1.0\linewidth]{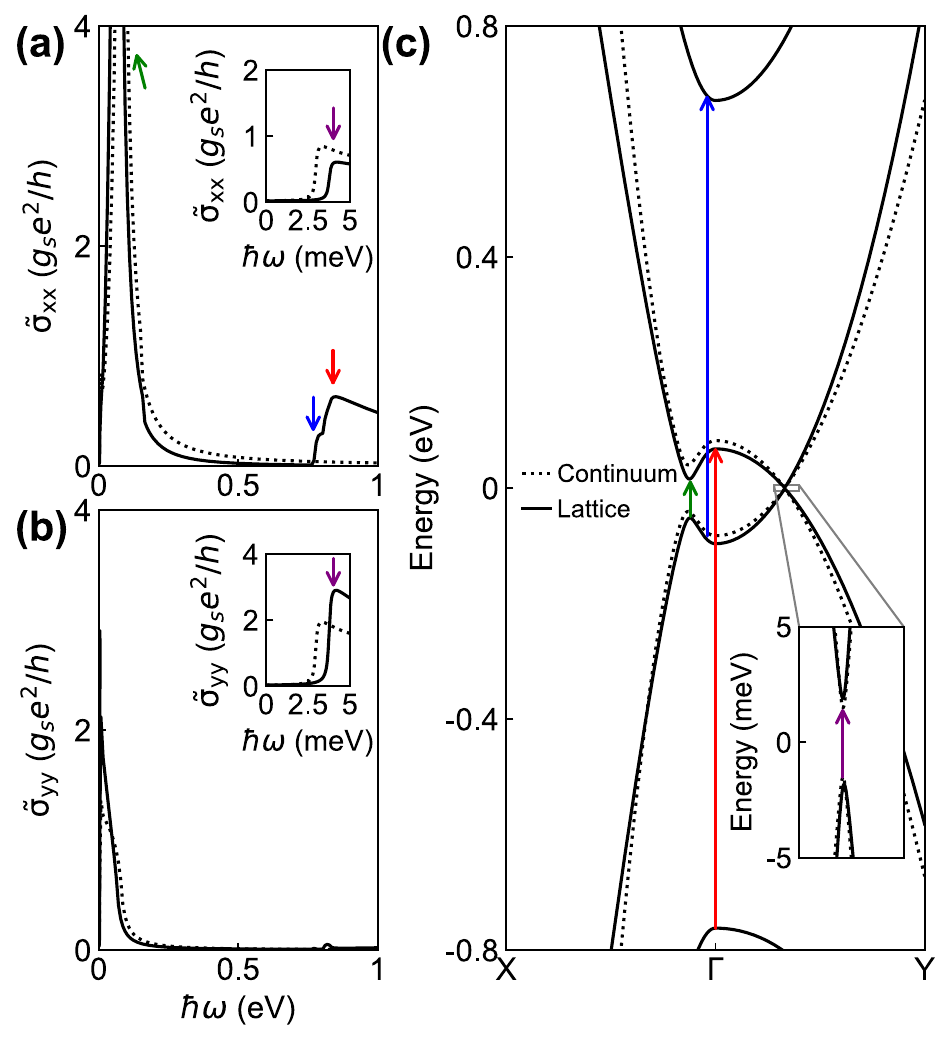}
\caption{
Longitudinal conductivities  (a) $\tilde{\sigma}_{xx}(\omega)$ and (b) $\tilde{\sigma}_{yy}(\omega)$ as a function of $\omega$, and (c) the band structure of tetralayer BP for $E_{\rm{ext}} = 0.5$ V/\r{A} and $\Lambda = 0.2$ \r{A}$^{-1}$. The results are presented for both the lattice model (solid line) and the continuum model (dotted line). The insets highlight the optical transitions across the Floquet gap $E_{\rm{FG}}$. The arrows in (c) mark interband transitions associated with the kink features in $\tilde{\sigma}_{\mu\mu}(\omega)$.
} 
\label{fig:fig8}
\end{figure}

\section{Longitudinal conductivity}

Figure~\ref{fig:fig8} presents the longitudinal conductivity of tetralayer BP in the Dirac semimetal phase for $\Lambda= 0.2$ \r{A}$^{-1}$. When $\varepsilon_{\rm {\Gamma}}$ is negative, two nodal points exist along the zigzag (Y) direction, and the Floquet gap $E_{\rm{FG}}$ opens upon applying time-periodic circularly polarized light $\bm{A}(t) = A_0 (\cos{\Omega t},\sin{\Omega t})^T$. Thus, the low-frequency optical conductivity can be interpreted as the sum of contributions from the two independent 2D gapped Dirac nodes with anisotropic in-plane velocities $v_x\approx v$ and $v_y\approx v_{\rm{D}}$, and a gap $E_{\rm{FG}}=2\Delta^2/\hbar\Omega$, where $\Delta=\hbar\sqrt{v_x v_y}\Lambda$ and 
$\Lambda={e A_0\over\hbar c}$. The analytical expression for the real part of the longitudinal conductivity of a single gapped Dirac node in Eq.~(\ref{eq: HFeff near Dirac nodes}) including the spin degeneracy factor $g_{\rm s}=2$ is given by \cite{Shao2019, Jeon2023}

\begin{equation}
    \!\!\sigma_{\mu\mu}(\omega)\! =\! \frac{g_{\rm s} e^2}{16 \hbar}\frac{v_{\mu}^2}{v_x v_y}\bigg[1 + \bigg(\frac{E_{\rm{FG}}}{\hbar\omega}\bigg)^2\bigg]\Theta(\hbar\omega - E_{\rm{FG}}).
\end{equation}

\noindent The first peak appears around $\hbar\omega=E_{\rm{FG}}\approx 4$ meV, as indicated by the purple arrow in Fig.~\ref{fig:fig8}.

The higher-energy optical features resemble those observed in nonirradiated tetralayer BP. In the high-energy limit, the Floquet eigenstates exhibit properties of the nondriven Hamiltonian~\cite{Bukov2015}, which possesses mirror symmetry $\mathcal{M}_y$ with respect to the $y = 0$ plane. As a consequence of the corresponding selection rules, $\tilde{\sigma}_{xx}$ shows distinct optical peaks, as indicated by the green, blue, and red arrows in Fig.~\ref{fig:fig8}, whereas these peaks are suppressed in $\tilde{\sigma}_{yy}$, consistent with the behavior in the absence of irradiation~\cite{Jang2019}.

\section{Light-induced Floquet gaps under elliptical polarization}\label{sec: elliptical polarization}

\begin{figure}[htb]
\includegraphics[width=1.0\linewidth]{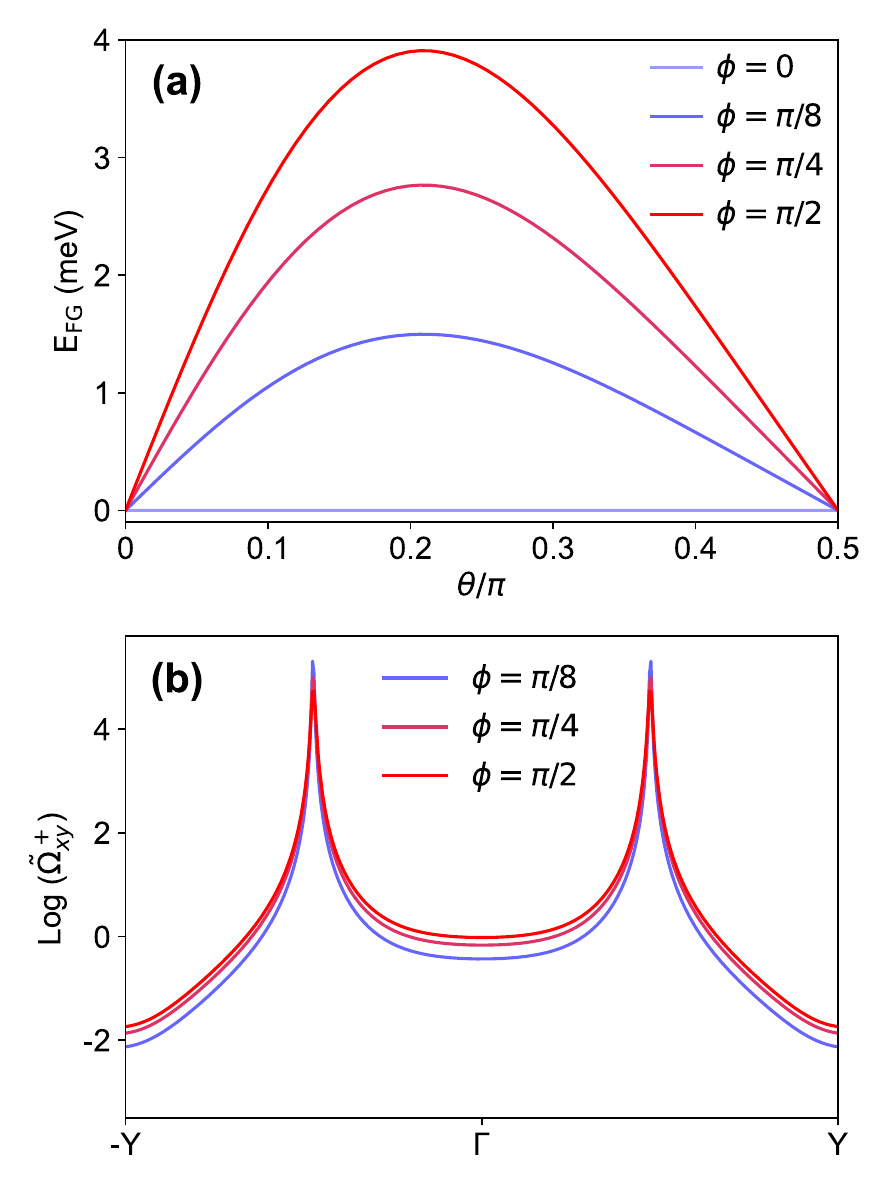}
\caption{
(a) The Floquet gap $E_{\rm{FG}}$ in the Dirac semimetal phase as a function of $\theta/\pi$ for $\phi = 0, \frac{\pi}{8}, \frac{\pi}{4},$ and $\frac{\pi}{2}$. (b) Berry curvature in the Dirac semimetal phase with $A_x = A_y$ ($\theta = \frac{\pi}{2}$) for $\phi = \frac{\pi}{8}, \frac{\pi}{4}$, and $\frac{\pi}{2}$. Note that for $\phi=0$, the Berry curvature vanishes due to the presence of time-reversal and inversion symmetries. For both (a) and (b), we used the light amplitude $\Lambda = \sqrt{\Lambda_x^2 + \Lambda_y^2} = 0.2$ \r{A}$^{-1}$ and external electric field $E_{\rm{ext}} = 0.5$ V/\r{A}.}
\label{fig:fig9}
\end{figure}

Given the intrinsic anisotropy of the structure and low-energy Hamiltonian of fBP, we extend our analysis to the case of time-periodic driving by elliptically polarized light. The vector potential is expressed as $\bm{A}(t) =  [A_x\cos{\Omega t},A_y\cos{(\Omega t-\phi)}]^T$, which is known as the Jones vector \cite{Valagiannopoulos2017, Seshadri2022, Orazbay2024}. Here, $A_x = A_0 \cos{\theta}$ and $A_y = A_0\sin{\theta}$ represent the polarization components along the armchair and zigzag directions, respectively, with the phase difference $\phi$. Note that the case with $A_x = A_y$ and $\phi =\frac{\pi}{2}$ ($\phi = -\frac{\pi}{2}$) corresponds to right- (left-) circularly polarized light. Under such periodic driving, the low-energy continuum Hamiltonian is modified via minimal coupling $\bm{k} \rightarrow \bm{k} + \frac{e}{\hbar c}\bm{A}(t)$, and the resulting effective Floquet Hamiltonian becomes
\begin{equation}\label{elliptical dirac continuum}
    H_{\rm{eff}}^{\rm{F}} \approx \hbar vq_{x}\sigma_{y} + \hbar v_{D} q_{y}\tau_{z}\sigma_{z} - \frac{\Delta_{xy}^{2}}{\hbar \Omega} \sin{\phi}\tau_{z}\sigma_{x},
\end{equation}
where $\Delta_{xy} = \hbar\sqrt{vv_{D}\Lambda_{x}\Lambda_{y}}$ with $\Lambda_{i}=\frac{eA_{i}}{\hbar c}$. This means that the magnitude of Floquet energy gap $E_{\rm{FG}}=2\left|\frac{\Delta_{xy}^2}{\hbar\Omega}\sin\phi\right|$ depends on the ellipticity parameters $A_{x}, A_{y},$ and $\phi$, and the Dirac cones becomes gapless for linearly polarized light, where $A_x=0$ or $A_y=0$ or $\sin{\phi} = 0$. In our framework, the effect of light helicity can be captured through the phase difference $\phi$. For circularly polarized light, changing $\phi$ to $\phi + \pi$ reverses helicity, which changes the sign of Berry curvature and dc Hall conductivity. 

We also performed a self-consistent lattice Hartree calculation in the presence of a static electric field under elliptical polarization of light. As shown in Fig.~\ref{fig:fig9}, the numerical calculations qualitatively agree with the continuum model in Eq.~(\ref{elliptical dirac continuum}). Both the continuum and lattice models show that $E_{\rm{FG}}$ is maximized for the circularly polarized light and becomes zero for the linearly polarized light. However, the lattice model yields asymmetric behavior in terms of $A_x$ and $A_y$ unlike the continuum model. This discrepancy arises because Eq.~(\ref{elliptical dirac continuum}) involves only the leading-order terms of the lattice model at the Dirac cones. The presence of the higher-order terms (such as the term proportional to $k_xk_y^2 \sigma_y$ as in Ref. \cite{Kim2017}) naturally makes it asymmetric. Even so, the Chern number remains unchanged as long as the helicity of the light is preserved, resulting in a quantized Hall conductivity.

\end{document}